\documentclass[%
 reprint,
superscriptaddress,
nofootinbib,
 amsmath,amssymb
 aps,
]{revtex4-1}

\usepackage{amsmath, amssymb}	% Loads AMS packages
\usepackage{bbm}
\usepackage{bm}				% Allows for bold Greek characters
\usepackage{braket}				% Allows Dirac notation
\usepackage{graphicx} 			% Allows for insertion of figures
\usepackage[usenames, dvipsnames]{color}		% Allows for text colour
\usepackage{mathtools}			% Use for small matrices
\usepackage{suffix}				% Allows for defining * version of custom commands
\usepackage{pst-math, pst-plot}	% Allows for post script pictures
\usepackage{mathtools}
\usepackage[hidelinks]{hyperref}
\usepackage{cleveref}
\usepackage{colonequals} %Get a nice looking \ce
\newcommand{\ce}{\colonequals}

\newcommand{\abs}[1]{\left| #1 \right|} 	% for absolute value
\DeclareMathOperator{\tr}{tr} 			% for trace

\newcommand{\nn}{\nonumber}		% for no numbering

\usepackage{soul}

\begin{document}

\title{Communicating without shared reference frames}

\author{Alexander R. H. Smith}
\email[]{alexander.r.smith@dartmouth.edu}
\affiliation{Department of Physics and Astronomy, Dartmouth College, Hanover, New Hampshire 03755, USA}

\date{\today}	% It is always \today, today,
	             	% but any date may be explicitly specified

\begin{abstract}
We generalize a quantum communication protocol introduced by Bartlett \emph{et al.} [New. J. Phys. 11, 063013 (2009)] in which two parties communicating do not share a classical reference frame, to the case where changes of their reference frames form a one-dimensional noncompact Lie group. Alice sends to Bob the state $\rho_R \otimes \rho_S$, where $\rho_S$ is the state of the system Alice wishes to communicate and $\rho_R$ is the state of an ancillary system serving as a token of her reference frame. Because Bob is ignorant of the relationship between his reference frame and Alice's, he will describe the state $\rho_R \otimes \rho_S$ as an average over all possible reference frames. Bob measures the reference token and applies a correction to the system Alice wished to communicate conditioned on the outcome of the measurement. The recovered state $\rho_S'$ is decohered with respect to $\rho_S$, the amount of decoherence depending on the properties of the reference token $\rho_R$. We present an example of this protocol when Alice and Bob do not share a reference frame associated with the one-dimensional translation group and use the fidelity between $\rho_S$ and $\rho_S'$ to quantify the success of the recovery operation.
\end{abstract}

%\pacs{Valid PACS appear here}% PACS, the Physics and Astronomy
                             % Classification Scheme.
%\keywords{Suggested keywords}%Use showkeys class option if keyword
                              %display desired
\maketitle

\section{Introduction}
\label{Introduction}

Most quantum communication protocols assume that the parties communicating share a classical background reference frame. For example, suppose Alice wishes to communicate to Bob the state of a qubit using a teleportation protocol \cite{Nielsen:2010}. Alice begins by having the qubit she wishes to communicate to Bob interact with one half of an entangled pair of qubits shared by her and Bob. Alice then measures the two qubits in her possession and picks up the phone and informs Bob of the measurement result. Bob uses this information to apply an appropriate gate to his half of the entangled pair to recover the state Alice wished to send to him. 

The success of this protocol depends on Alice's ability to classically communicate to Bob which gates he should apply to his half of the entangled state. This can only be done if Alice and Bob share a reference frame. As an example, suppose Alice informs Bob that he needs to apply the Pauli $z$ operator to the qubit in his position. If Bob is ignorant of the orientation of his lab with respect to Alice's, he does not know in which direction to orient the magnetic field in his Stern-Gerlach apparatus to implement the Pauli $z$ operator to recover the state sent by Alice. In this case the teleportation protocol is unable to be carried out perfectly \cite{Giulio-Chiribella:2012, Verdon:2018,*Verdon2:2018}.

This motivates the study of quantum communication without a shared reference frame \cite{Bartlett:2007}. One way Alice can communicate to Bob, despite not sharing a reference frame with him, is to encode information into degrees of freedom that are invariant under a change of Alice's reference frame. Without knowing his relation to Alice's reference frame, Bob is able to extract both classical and quantum information encoded in these degrees of freedom \cite{Rudolph:2003}. However, in practice such communication schemes may be challenging to implement since they require highly entangled states of many qubits.

Another possibility for Alice and Bob to communicate without a shared reference frame is for Alice to send Bob a quantum system $\rho_R$ to serve as a token of her reference frame, together with the state $\rho_S$ she wishes to communicate to him. Since Bob does not know the relation between his reference frame and Alice's, with respect to his reference frame he will see the joint state $\rho_R \otimes \rho_S$ averaged over all possible orientations of his lab with respect to Alice's; this averaging operation is referred to as the \mbox{$G$-twirl} and the averaged state denoted as $\mathcal{G} [\rho_R \otimes \rho_S]$. Bob can apply a recovery operation to this $G$-twirled state by measuring the reference token and applying an appropriate correction to the system Alice wishes to send to him, allowing him to recover a state $\rho_S'$ that is close to $\rho_S$. This recovery operation was first constructed by Bartlett \emph{et al.} \cite{Bartlett:2009}, and its success was found to depend on the size of the reference token, which is necessarily bounded if the reference token is described by a finite dimensional Hilbert space.

However, this communication protocol is based on Bob assigning the $G$-twirled state $\mathcal{G} [\rho_R \otimes \rho_S]$ to the system and reference token, and the $G$-twirl does not yield normalizable states when the group of reference frames being averaged over is noncompact \cite{Smith:2016}. This begs the question: Can an analogous communication protocol involving a reference token sent by Alice and a recovery operation implemented by Bob be constructed given that changes of their reference frames form a noncompact group? Furthermore, if the Hilbert space of the reference token is infinite dimensional, for example $\mathcal{H}_R \simeq L^2(\mathbb{R})$, what physical aspect of the reference token acts as its effective size?

The purpose of this article is to examine these questions. Considerations of noncompact groups within the theory of quantum reference frames is important if one hopes to apply the theory to the physically relevant Galilean and Poincar\'{e} groups, which are both noncompact.

We begin in Sec.~\ref{Communication with our a shared reference frame} by describing the encoding and recovery operations introduced by Bartlett \emph{et al.} \cite{Bartlett:2009}. In Sec.~\ref{A recovery operation for noncompact groups} we introduce a $G$-twirl over a compact subset of a noncompact group and a complementary recovery operation, such that in the limit when this $G$-twirl becomes an average over the entire noncompact group, the composition of the recovery operation with this $G$-twirl results in properly normalized states. We then apply this construction in Sec.~\ref{Application to reference frames with the translation group} to the case when Alice and Bob do not share a reference frame associated with the one-dimensional translation group, which is relevant for parties communicating without a shared positional reference frame. In this case, we identify the inverse of the width in position space of the reference token's state as the effective size of the reference token and demonstrate that in the limit when this width goes to zero Alice and Bob are able to communicate perfectly without a shared reference frame. We conclude in Sec.~\ref{SummaryCh7} with a summary of our results and an outlook to future questions.

%========================================
%========================================

\section{Communication without a shared classical reference frame}
\label{Communication with our a shared reference frame}

Consider two parties, Alice and Bob, each employing their own classical reference frame to describe the state of a single quantum system associated with the Hilbert space $\mathcal{H}_S$. Suppose that this system transforms via a unitary representation of the group $G$ when changing the reference frame used to describe the system; for the time being we will assume $G$ is a compact Lie group. 

Let $g\in G$ label the group element which describes the transformation from Alice's to Bob's reference frame. If Alice prepares the system in the state $\rho_S \in \mathcal{S}(\mathcal{H}_S)$ with respect to her reference frame, where $\mathcal{S}(\mathcal{H}_S)$ is the space of states on $\mathcal{H}_S$, and $g$ is completely unknown to Bob, then the state with respect to his reference frame will be given by a uniform average over all possible $g \in G$; that is, by the $G$-twirl
\begin{align}
\mathcal{G}[\rho_S] \ce \int_G d g \, U_S(g) \, \rho_S \, U_S(g)^\dagger, \label{firstGtwirl}
\end{align}
where $d g$ denotes the Haar measure associated with $G$ and $U_S(g)\in \mathcal{U}(\mathcal{H}_S)$ is the unitary representation of the group element $g\in G$ on $\mathcal{H}_S$, with $\mathcal{U}(\mathcal{H}_S)$ denoting the space of unitary operators on $\mathcal{H}_S$. If instead Bob has some partial information about the relation between his reference frame and Alice's, the uniform average over all possible $g\in G$ in Eq.~\eqref{firstGtwirl} would be replaced with a weighted average encoding Bob's partial information~\cite{Miatto:2012, Ahmadi:2015}.

In general, the $G$-twirl results in decoherence, not from the system interacting with an environment and information being lost to the environment, but from Bob's lack of knowledge about the relationship between his reference frame and Alice's. To combat this decoherence, Alice may prepare another quantum system, described by the Hilbert space $\mathcal{H}_R$, to serve as a token of her reference frame (a good representative of her reference frame). Suppose Alice prepares the token in the state $\ket{e} \in \mathcal{H}_R$, then the reference token and system relative to Bob's frame will be given by the encoding operation
\begin{align}
\mathcal{E}[\rho_S] &\ce \mathcal{G} \big[ \ket{e} \! \bra{e} \otimes \rho_S \big] \nn \\
&= \int_G dg \,  \mathcal{U}_R(g)\!\left[ \ket{e}\! \bra{e} \right] \otimes  \mathcal{U}_S(g)\!\left[ \rho_S \right],
\label{encoding}
\end{align}
where $\mathcal{U}_i(g) \! \left[ \rho \right]  \ce  U_i(g) \, \rho  \, U_i(g)^\dagger$ denotes the adjoint representation of the action of the group element $g\in G$ on $\rho \in \mathcal{S} \left(\mathcal{H}_i \right)$ for  $i \in \{R, S\}$.

Bob's task is now to best recover the state of the system $\rho_S$ given the encoded state $\mathcal{E}[\rho_S]$. In other words, he must construct a recovery operation 
\begin{align}
\mathcal{R}: \mathcal{S}(\mathcal{H}_R\otimes \mathcal{H}_S) \to \mathcal{S}(\mathcal{H}_S),
\end{align}
that when applied to $\mathcal{E}[\rho_S]$  results in a state $\rho_S' \in \mathcal{S}(\mathcal{H_S})$ that is as close as possible to $\rho_S$. A recovery operation %\footnote{The Hilbert-Schmidt adjoint of $\mathcal{E}$.}  
$\mathcal{R}$ was constructed by Bartlett \emph{et al.} \cite{Bartlett:2009} with such properties, and its action on the encoded state $\mathcal{E}[\rho_S]$ yields
\begin{align}
\rho_S' \ce \mathcal{R} \circ \mathcal{E}[\rho_S] = \int_G d g \, p\!\left(g\right) \mathcal{U}_S(g) \!\left[  \rho_S \right],
\label{composition}
\end{align}
where $p\!\left(g\right) \propto \abs{\braket{e| U_R(g) | e}}^2$ with $U_R(g)\in \mathcal{U}(\mathcal{H}_R)$ being the unitary representation of $g\in G$ on $\mathcal{H}_R$. We will explicitly construct a similar recovery operation in the next section for the case when $G$ is noncompact.

\begin{figure}[h]
\includegraphics[width =0.45\textwidth]{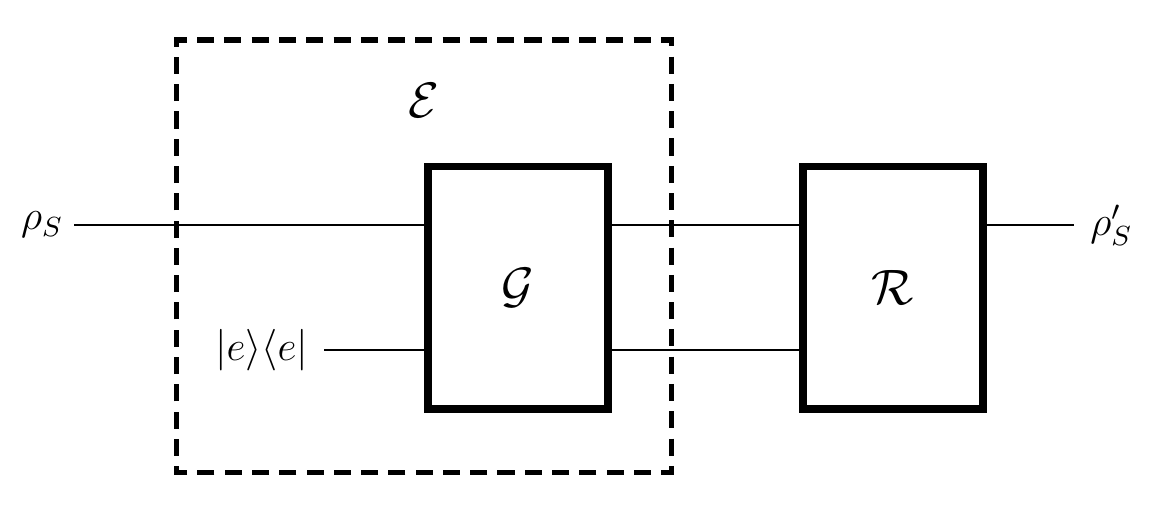}
\caption{
The communication channel $\mathcal{R} \circ \mathcal{E}$. Alice prepares a state $\rho_S$ she wishes to communicate to Bob along with the state $\ket{e}\!\bra{e}$ as a token of her reference frame. As Bob does not know the relation between his reference frame and Alice's, he sees the joint state of the reference token and system as the encoded state $\mathcal{E} [\rho_S] = \mathcal{G} \big[ \ket{e} \! \bra{e} \otimes \rho_S \big]$. Bob then applies the recovery operation to the encoded state and recovers the state $\rho_S' = \mathcal{R} \circ \mathcal{E} [\rho_S]$.
 }
\label{CommunicationChannel}
\end{figure}

%========================================
\section{A recovery operation for noncompact groups}
\label{A recovery operation for noncompact groups}

The action of the $G$-twirl over a noncompact group on a state results is a non-normalizable density matrix \cite{Smith:2016}. For example, consider the $G$-twirl over the non-compact group of translations in one dimension $T_1$ of the state $\rho \in \mathcal{S} \left(  L^2(\mathbb{R})\right)$. The unitary representation of $g\in T_1$ is $U(g) = e^{-iPg}$, where $P$ is the momentum operator on $L^2(\mathbb{R})$, and the $G$-twirl over $T_1$ is
\begin{align}
\mathcal{G}_{T_1}\!\left[\rho \right] &= \int d{g} \,  e^{-i g  {P}}  \left(\int dp dp' \,  \rho\! \left(p, p' \right) \ket{p}\!\bra{p'} \right) e^{ig  {P}} \nn \\
& =  2\pi \int d{p} \, \rho \!\left(p,p \right) \ket{p}\!\bra{p}, \label{TranslationTwirl}
\end{align}
where $\ket{p}$ denote the eigenkets of the momentum operator $P$, $\rho\! \left(p, p' \right) \ce  \braket{p' | \rho | p}$, and $dg$ is the Haar measure\footnote{Even though $G$ is a noncompact Lie group, it is still locally compact, and thus possesses a nontrivial left invariant Haar measure that is unique up to a positive constant~\cite{Nachbin:1965}. In the case of the translations group considered here, $dg$ corresponds to the Lebesgue measure on the real line.} associated with $T_1$; in going from the first to the second equality we have used the definition of the Dirac delta function $2 \pi \delta(p-p') \ce  \int d g \, e^{ig(p-p')}$. From Eq.~\eqref{TranslationTwirl} it is clear that $\mathcal{G}_{T_1}\!\left[\rho \right] \notin \mathcal{S}\left( \mathcal{H} \right)$, which can be verified by computing the norm of $\mathcal{G}_{T_1}\!\left[\rho \right]$ which is infinite.

Given that the codomain of the $G$-twirl over a noncompact group does not necessarily correspond to the state space $\mathcal{S}(\mathcal{H})$, it is not clear whether the encoding operation $\mathcal{E}$ or the recovery operation $\mathcal{R}$ discussed above are applicable to reference frames associated with noncompact groups. We now demonstrate that despite this fact, the composition of an encoding operation associated with a noncompact group with a suitably defined recovery operation results in a properly normalized state.

The approach we will take is to define a compact $G$-twirl over a compact subset of the noncompact group $G$ associated with the reference frame, which corresponds to Bob having partial information that the relation between his reference frame and Alice's is described by $g \in [-\tau, \tau] \subset G$. This compact $G$-twirl will be used in an encoding operation analogous to Eq.~\eqref{encoding}.  We will then construct a complementary recovery operation, compose it with this encoding operation (similar to Eq.~\eqref{composition}), and finally take a limit in which the compact $G$-twirl corresponds to a uniform average over the entire noncompact group $G$. We will show that in this limit the recovered state is properly normalized and contained in $\mathcal{S} ( \mathcal{H}_S )$.

%========================================
\subsubsection{The encoding map}

Consider all possible transformations of Alice's and Bob's classical reference frames to form a strongly continuous one-parameter noncompact Lie group $G$. Suppose that the unitary representation of a group element $g\in G$ on the Hilbert space $\mathcal{H}_R$ describing the reference token is $U_R(g)\in \mathcal{U}(\mathcal{H}_R)$. By Stone's theorem~\cite{Stone:1930}, $U_R(g) = e^{igA_R}$ is generated by a self-adjoint operator ${A}_R$, the spectrum of which we denote by $\sigma(A_R)$ and assume to be continuous\footnote{This is true of the group generated by either the position or momentum operator on $L^2(\mathbb{R})$. We note that the following construction does not rely on $\sigma(A_R)$ being continuous. }. For each element of the spectrum $f(a_R) \in \sigma(A_R)$ there corresponds an eigenket $\ket{a_R}$ such that
\begin{align}
{A}_R \ket{a_R} =f(a_R) \ket{a_R},
\label{eigenEquation}
\end{align}
with eigenvalue $f(a_R) \in \mathbb{R}$. Since $\sigma(A_R)$ is continuous and $A_R$ is self-adjoint, these eigenkets  are normalized with the Dirac delta function
\begin{align}
\braket{a_R|a_R'} = \delta\big(a_R-a_R'\big).
\end{align}
From the above normalization condition we see that $\ket{a_R} \not\in \mathcal{H}_R$, as these eigenkets are not square integrable and therefore do not represent physical states\footnote{More precisely~\cite{Ballentine:1998}, when dealing with operators with continuous spectrum the theory is defined on a rigged Hilbert space defined by the triplet $\Phi \subset \mathcal{H}_R \subset  \Phi'$, where $\Phi$ is a proper subset dense in $\mathcal{H}_R$ and $\Phi'$ is the dual of $\Phi$, defined through the inner product on $\mathcal{H}_R$. In our case, $\Phi$ is the Schwarz space of smooth rapidly decreasing functions on $\mathbb{R}$ and $\Phi'$ is the space of tempered distributions on $\mathbb{R}$. The eigenkets $\ket{a_R}$ are in $\Phi'$.}.

Our first step is to construct a well defined encoding operation analogous to Eq.~\eqref{encoding}. To do so, we suppose the state of Alice's reference token $\ket{e} \in \mathcal{H}_R$, expressed in the basis furnished by the eigenkets of ${A}_R$, is
\begin{align}
\ket{e} \ce \int d a_R \, \psi_R(a_R) \ket{a_R},\label{ReferenceTokenState}
\end{align}
where $\psi_R(a_R) \ce \braket{a_R | e}$. Next, let us introduce the set of states
\begin{align}
\big\{ \ket{e(g)} \ce U_R(g) \ket{e}  \, \big| \, \forall g \in G \big\},
\label{SetOfReferenceTokens}
\end{align}
where each $\ket{e(g)}$ corresponds to a different orientation of Alice's reference frame. The state of the reference token $\ket{e}$ should be chosen such that each $\ket{e(g)}$ defined in Eq.~\eqref{SetOfReferenceTokens} is distinct, that is, the state of the reference token should not be invariant with respect to $G$. Furthermore, for the states $\ket{e(g)}$ to imitate a classical reference frame, they must be orthogonal
%\begin{align}
%\braket{g|g'} = \delta_{g,g'}, \label{critera1}
%\end{align} 
so as they are perfectly distinguishable.
%; we will refer to these such states as ideal states of the reference token. However, it may be that no ideal state of the reference token exists in $\mathcal{S}( \mathcal{H}_R )$, such is the case the reference tokens associated with the translation group studied in the following section. In this case, the state of the reference token should be chosen from a sequence of functions $\psi_\sigma(a_R)$ such that $ \lim_{\sigma \to 0}\abs{\psi_\sigma(a_R)}^2 = \abs{\psi(a_R)}^2$. %Note that Eq.~\eqref{critera3} only determines the ideal reference token state up to a phase $e^{i \phi(a_R) }$ where $\phi(a_R) \in \mathbb{R}$.

%If Alice prepared the reference token in the state $\ket{e_\sigma} \ce \int d a_R \, \psi_\sigma(a_R) \ket{a_R} \in \mathcal{H}_R$, the states of the reference token corresponding to different orientations of Alice's reference frame are
%\begin{align}
%\ket{g_\sigma}= U_R(g) \ket{e_\sigma} =  \int d a_R \, \psi_\sigma(a_R) e^{i f(a_R) g}  \ket{a_R }.
%\label{defg}
%\end{align}
%%\tcr{[Do these states furnish a basis for $\mathcal{H}_R$?]}

Now suppose Alice prepares her reference token in the state $\rho_R  \in \mathcal{S}(\mathcal{H}_R)$
%\begin{align}
%\rho_R = \ket{e}\!\bra{e} \in \mathcal{S}(\mathcal{H}_R), \label{referenceToken}
%\end{align}
%, where
%\begin{align}
%\ket{e_\sigma} \ce \int d a_R \, \psi_\sigma(a_R) \ket{a_R} \in \mathcal{H}_R,
%\end{align}
and wishes to send Bob the state $\rho_S \in \mathcal{S}(\mathcal{H}_S)$ of a system associated with the Hilbert space $\mathcal{H}_S$. If Bob knows the relation between his reference frame and Alice's is given by a group element $g \in [-\tau,\tau]  \subset G$, but within this interval he is completely ignorant of which group element corresponds to this relation, he will describe the joint state of the reference token and system by the output of the encoding operation
\begin{align}
\mathcal{E}_\tau :  \   \mathcal{S}(\mathcal{H}_S) &\to \mathcal{S}(\mathcal{H}_R \otimes \mathcal{H_S}) \nn \\
\rho_S &\mapsto \mathcal{E}_\tau[\rho_S] \ce \mathcal{G}_\tau \big[ \rho_R \otimes \rho_S \big],
\label{tauEncoding}
\end{align}
where the map $\mathcal{G}_\tau$ is a uniform average of $\rho_S$ over the compact interval $[-\tau, \tau] \subset G$,
\begin{align}
\mathcal{G}_\tau \left[ \rho_R \otimes \rho_S \right]  \ce \frac{1}{2\tau} \int^\tau_{-\tau} d g \  \mathcal{U}_R(g) \! \left[ \rho_R \right] \otimes \mathcal{U}_S(g) \! \left[ \rho_S \right],
\end{align}
where $dg$ is the Haar measure associated with $G$.

%\begin{align}
%\mathcal{G}_\tau :  \ \mathcal{S}(\mathcal{H}_R \otimes \mathcal{H_S}) &\to \mathcal{S}(\mathcal{H}_R \otimes \mathcal{H_S}) \nn \\
%\ket{e} \! \bra{e} \otimes \rho_S &\mapsto \mathcal{G}_\tau \big[ \ket{e} \! \bra{e} \otimes \rho_S \big] \nn \\
%&\quad \ce \frac{1}{2\tau} \int^\tau_{-\tau} d g \  \mathcal{U}_R(g) \left[ \rho_R \right] \otimes \mathcal{U}_S(g) \left[ \rho_S \right] ,
%\end{align}
%and we have defined the maps $\mathcal{U}_i(g) \left[ \rho_i \right]  \ce  U_i(g) \, \rho_i  \, U_i(g)^\dagger$ for  $i \in \{R, S\}$.
%\begin{align}
%\mathcal{U}_R(g) \left[ \rho_R \right]  &\ce  U_R(g) \, \rho_R  \, U_R(g)^\dagger, \nn  \\
%\mathcal{U}_S(g) \left[ \rho_S \right] &\ce  U_S(g)  \, \rho_S  \, U_S(g)^\dagger. \nn
%\end{align}
%In the limit $\tau$ is taken to infinity, the map $\mathcal{G}_\tau$ converges to the $G$-twirl over the entire group $G$. %\tcr{\bf [We can be vague and not specify the operator topology we are using define this convergence. However, if I understood our past discussion correctly, in some sense we are using the fidelity as a norm. Let's discuss. Also, if we can use the fidelity as a `norm', I think that Fig. 2 shows that the convergence is not uniform. Although, I] }

%========================================
\subsubsection{The recovery operation}

As demonstrated by Bartlett \emph{et al.} \cite{Bartlett:2009}, Bob may perform a recovery operation $\mathcal{R}$ by first making a measurement of the reference token, followed by a reorientation of the system conditioned on the outcome of the measurement, and then discarding both the reference token and measurement result. We follow this procedure in constructing the recovery operation to be applied to the encoded state $\mathcal{E}_\tau(\rho_S)$.

%Assume there exists a self-adjoint operator $B_R$ acting on $\mathcal{H}_R$ which is canonically conjugate to the generator $A_R$, that is, $[A_R, B_R] =i$. Let $P_g \in \mathcal{E}(\mathcal{H})$ denote the effect operator associated with 

Bob will make a measurement $R$ of the reference token described by the POVM elements  
\begin{align}
R \ce \big\{ dg \, E(g) ,  \ \forall g \in [-\tau, \tau] \subset G\big\} \cup \big\{ {E}_\tau \big\},
\end{align}
where
\begin{align}
{E}_\tau \ce I_R -\int_{-\tau}^\tau d g \,E(g),
\label{Edef}
\end{align}
$dg \, E(g)$ is the POVM element associated with outcome $g \in G$, and $I_R$ is the identity operator on $\mathcal{H}_R$. We assume\footnote{To the best of the authors' knowledge the question of whether such a measurement exists for any $G$ is an open problem, as suggested by the remarks in Sec. III.4.4 of Ref. \cite{Busch:1997}. However, it is suggested in this reference that it seems plausible that such a measurement can be constructed, although there does not seem to be an easy general procedure for its construction. Nonetheless, such measurements exist for physically relevant groups like the translation group considered in the following section.} that these POVM elements satisfy the covariance relation
\begin{align}
\mathcal{U}_R(g')\!\left[E(g)\right]  = E(g + g') \quad \forall g \in G. \label{covariance}
\end{align}

%, and $\ket{g} = \lim_{\sigma \to 0} U(g) \ket{e_\sigma}$ \tcr{\bf [Maybe comment on additional error if Bob isn't able to actually make measurements of $\ket{g}$ in the limit $\sigma \to 0$?]}. In the limit when $\tau$ becomes infinite, the measurement of $R$ reduces to the one considered in~\cite{Bartlett:2009}.

If the outcome of the measurement of ${R}$ is $g \in [-\tau,\tau] $, associated with the POVM element $ dg \, E(g)$, then Bob will reorient the system by implementing the unitary map $\mathcal{U}_{S}(g^{-1})$, which corresponds to the transformation of the reference token by an amount indicated by the measurement result (1st term in Eq.~\eqref{rhoStau}). If the outcome of the measurement is associated with the operator ${E}_\tau$,  Bob will do nothing (2nd term in Eq.~\eqref{rhoStau}). After this measurement and reorientation, Bob will discard (trace out) the reference token and measurement result. This entire procedure will constitute the recovery operation $\mathcal{R}_\tau$.

The action of the recovery operation $\mathcal{R}_\tau$ on the encoded state $\mathcal{E}_\tau[\rho_S]$ is given by
\begin{widetext}
\begin{align}
\rho_S'(\tau) &= \mathcal{R}_\tau \circ \mathcal{E}_\tau[\rho_S] \nn \\
&= \frac{1}{2\tau} \int_{-\tau}^\tau d g' \int_{-\tau}^\tau  d g  \,  \tr \big( E\!\left(g'\right)  \mathcal{U}_R(g) \left[ \rho_R \right] \big) \, \mathcal{U}_S(g'^{-1}) \circ \mathcal{U}_S(g) \left[ \rho_S \right]   +\frac{1}{2\tau} \int_{-\tau}^\tau d g  \, \tr \big( {E}_\tau \, \mathcal{U}_R(g) \left[ \rho_R \right] \big) \, \mathcal{U}_S(g) \left[ \rho_S \right].
\label{rhoStau}
\end{align}
\end{widetext}

%========================================
\subsubsection{Taking the limit $\tau \to \infty$}
The limit of Eq.~\eqref{rhoStau} in which $\tau$ becomes infinite corresponds to the scenario in which Bob knows nothing about the orientation of his reference frame with respect to Alice's\,---\,the $G$-twirl appearing in the encoding map in Eq.~\eqref{tauEncoding} is an average over the entire group $G$.

As is clear from Eq.~\eqref{Edef}, in the limit $\tau \to \infty$ the operator $E_\tau$ vanishes, and thus the second term in Eq.~\eqref{rhoStau} goes to zero. Taking this into account,  the $\tau\to\infty$ limit of Eq.~\eqref{rhoStau} is
\begin{align}
\rho_S' &= \lim_{\tau\to \infty} \frac{1}{2\tau} \int_{-\tau}^\tau d g' \int_{-\tau}^\tau  d g  \, \tr \big( E(g' - g)  \rho_R \big) \nn \\
&\qquad \qquad \qquad \qquad \qquad \qquad  \times \mathcal{U}_S(g-g') \! \left[ \rho_S \right],
\label{rhoStau2}
\end{align}
where we have used the covariance property of the POVM elements expressed in Eq.~\eqref{covariance}.
Changing the integration variables to $u \ce g' - g$ and $v\ce g'$, the recovered state becomes
\begin{align}
\rho_S' &= \lim_{\tau\to \infty} \frac{1}{2\tau} \int_{-\tau}^\tau d v \int_{v-\tau}^{v+\tau}  d u  \, \tr \big(  E(u) \rho_R  \big) \,  \mathcal{U}^\dagger_S(u) \! \left[ \rho_S \right].
\label{rhoStau3}
\end{align}
Denoting the antiderivative of the above integrand as
\begin{align}
F(x) \ce \int_0^x du \,\tr \big(  E(u) \rho_R  \big) \,  \mathcal{U}^\dagger_S(u) \! \left[ \rho_S \right],
\end{align}
Eq.~\eqref{rhoStau3} takes the form
\begin{align}
\rho_S' &=  \lim_{\tau\to \infty} \frac{1}{2\tau} \int_{-\tau}^\tau d v \,\big(F(v+\tau) -F(v-\tau) \big).
\end{align}
Making the substitution  $h\ce\tau+v$ and $h\ce\tau-v$ in the first and second terms, respectively, the recovered state simplifies to
\begin{align}
\rho_S' &=  \lim_{\tau\to \infty} \frac{1}{2\tau} \int_{0}^{2\tau} d h \, \big(F(h) -F(-h) \big).
\end{align}
Taking the limit by applying L'H\^{o}pital's rule\footnote{Suppose $f(x)$ and $g(x)$ are real differentiable function in \mbox{$(a,b) \subset \mathbb{R}$}, and $g'(x) \neq 0$ for all $x \in (a,b)$. Further, suppose that $f'(x)/g'(x) \to A$ as $x \to a$. Then L'H\^{o}pital's rule states that if $f(x) \to 0$ and $g(x) \to 0$ as $x \to a$ or if $g(x) \to \infty$ as $x \to a$, then $f(x) / g(x) \to A$ as $x \to a$ \cite{Rudin:1976}.} yields 
\begin{align}
\rho_S'  &= \frac{1}{2} \lim_{\tau \to \infty} \frac{\partial}{\partial \tau}  \int_{0}^{2\tau} d h \, \big(F(h) -F(-h) \big) \nn \\
&=  \lim_{\tau \to \infty} \big(F(\tau) -F(-\tau) \big) \nn \\
&=   \int_G d g  \, \tr \big(  E(g) \rho_R  \big) \, \mathcal{U}_S(g) \! \left[ \rho_S \right],
\end{align}
where the integration is carried out over the entire group~$G$.

This brings us to our main result: even though the action of the $G$-twirl over a noncompact group yields non-normalizable states, the composition of the encoding operation, which makes use of the $G$-twirl, with the recovery operation applied to $\rho_S$ results in a properly normalized state in $\mathcal{S}(\mathcal{H}_S)$. Explicitly 
\begin{align}
\rho_S' &=  \lim_{\tau\to\infty}  \mathcal{R}_\tau \circ \mathcal{E}_\tau [\rho_S] 
\nn \\
&= \int_G d g \, p\!\left(g\right) \mathcal{U}_S(g) \! \left[ \rho_S \right] \in \mathcal{S}(\mathcal{H}_S),
\label{LimitOfCompostion}
\end{align}
%where
%\begin{align}
%p\!\left(g\right) \ce \frac{\abs{\braket{e|U(g)|e}}^2}{\int d g \, \abs{\braket{e|U(g)|e}}^2},
%\end{align}
where $p\!\left(g\right)\ce \tr \big(  E(g) \rho_R  \big)$ is a normalized probability distribution on $G$. %\tcr{Note that although we assumed  the reference token was prepared by Alice to be in a pure state (see Eq.~\eqref{referenceToken}), the limit expressed in Eq.~\eqref{LimitOfCompostion} still applies had the state of the reference token been mixed.}

Equation \eqref{LimitOfCompostion} is identical to the expression for the composition of the recovery and encoding map defined for compact groups given in Eq.~\eqref{composition}. From Eq. \eqref{LimitOfCompostion} we see that if $p\!\left(g\right)$ is highly peaked around the identity group element then the only unitary that will contribute significantly is the identity operator, and the state recovered by Bob will be close to the state sent by Alice, $\rho_S' \approx \rho_S$. Thus, the success of the recovery operation, and consequently the quality of the reference token, can be quantified in terms of the width of $p\!\left(g\right)$, analogous to the compact case \cite{Bartlett:2009}.

By expressing $\rho_S$  in the basis furnished by the eigenkets of the generator $A_S$ of the group $G$, we find the recovered state to be
\begin{align}
\rho_S' &=   \int_G dg \, p\!\left(g\right) \int da_S da_S' \, \rho_S(a_S,a_S') \nn \\
&\qquad \qquad \qquad \qquad \qquad \qquad  \times e^{i A_S g} \ket{a_S}\!\bra{a_S'} e^{-i A_S g} \nn \\
&= \int da_S da_S'  \, \left[ \int_G dg \, p\!\left(g\right)  e^{i g (a_S -a_S')} \right] \nn \\
&\qquad \qquad \qquad \qquad \qquad \qquad  \times\rho_S(a_S,a_S')  \ket{a_S}\!\bra{a_S'} \nn \\
&= \int da_S da_S' \, \tilde{p}(a_S - a_S')  \rho_S(a_S,a_S') \ket{a_S}\!\bra{a_S'},
\end{align}
where in the last equality we have defined the Fourier transform of $p\!\left(g\right)$
\begin{align}
\tilde{p}(a_S - a_S') \ce \int_G dg \, p\!\left(g\right)  e^{i g (a_S -a_S')}.
\end{align}
From the definition of the characteristic function \mbox{$\tilde{p}(a_S -a_S')$} above, we see that if $a_S = a_S'$,  then \mbox{$\tilde{p}(a_S - a_S') =1$}, and consequently the diagonal elements of $\rho_S$ are unaffected by the action of the communication channel $ \lim_{\tau\to\infty}  \mathcal{R}_\tau \circ \mathcal{E}_\tau$. On the other hand, since the characteristic function is bounded, $\abs{\tilde{p}(a_S - a_S') }\leq 1$, when $a_S \neq a_S'$ the off diagonal elements of $\rho_S'$ are equal to those of $\rho_S$ multiplied by a factor whose magnitude is less than or equal to unity. From this observation we see that the decoherence induced by  $\lim_{\tau\to\infty}  \mathcal{R}_\tau \circ \mathcal{E}_\tau$ occurs in the basis furnished by the eigenkets associated with the generator $A_S$ of the group $G$.

%In a differnt context, similar observations were made in \cite{Miatto:2012}.

To quantify the success of the recovery operation\,---\,how close the recovered state $\rho_S'$ is to the initial state $\rho_S$\,---\,we will make use of the fidelity $F(\rho_S', \rho_S)$ between the recovered state $\rho_S'$ and the state $\rho_S=\ket{\psi_S}\!\bra{\psi_S} \in \mathcal{S}(\mathcal{H}_R)$ that Alice sent, which we will take to be pure \begin{align}
\ket{\psi_S} = \int da_S \, \psi_S(a_S) \ket{a_S},
\end{align}
where $\psi_S(a_S) \ce \braket{a_S | e}$. The fidelity  $F(\rho_S',\rho_S)$ is then given by
\begin{align}
F(\rho_S',\rho_S) &\ce \braket{\psi_S | \rho_S' |\psi_S} \nn \\
&= \int_G dg \, p\!\left(g\right) \abs{\braket{\psi_S| U_S(g) | \psi_S } }^2 \nn \\
&= \int da_S da_S' \, \tilde{p}(a_S - a_S')  \abs{\psi_S(a_S)}^2 \abs{\psi_S(a_S')}^2. \label{FidelityResult}
\end{align}

%========================================
%========================================
\section{Reference frames associated with the translation group}
\label{Application to reference frames with the translation group}

We now examine the recovered state $ \rho_S'=  \lim_{\tau\to\infty}  \mathcal{R}_\tau \circ \mathcal{E}_\tau [\rho_S] $ when the relevant reference frame is associated with the one-dimensional translation group.

Consider Alice and Bob being completely ignorant of the relation between the spatial origins of their labs, i.e., the relation between their positional reference frames. The group formed by all possible changes of Alice's reference frame is the one-dimensional translation group $T_1$. The unitary representation of the group element $g\in T_1$ on the system is $U_S(g) \in \mathcal{U}_S ( \mathcal{H}_S)$ and on the reference token is $U_R(g) \in \mathcal{U}_R ( \mathcal{H}_R)$. These representations are generated by their respective momentum operators ${A}_S = {P}_S$ and ${A}_R ={P}_R$.

Suppose as a token of Alice's reference frame she prepares the state $\ket{e_\sigma} \in \mathcal{H}_R\simeq L^2(\mathbb{R})$, which we take to be a Gaussian state
\begin{align}
\ket{e_\sigma} &= \frac{1}{\pi^{1/4} \sqrt{\sigma} } \int dx_R \, e^{-  x_R^2/2\sigma^2}  \ket{x_R}, 
%\nn \\
%&= \frac{\sqrt{\sigma}}{\pi^{1/4}} \int dp_R \, e^{- \sigma^2 p_R^2/2}  \ket{p_R},
\end{align}
where we have expressed $\ket{e_\sigma}$ in the basis furnished by the eigenkets $\ket{x_R}$ of the position operator $X_R$ on $\mathcal{H}_R$ and $\sigma >0$ is the spread of this state with respect to this basis. Note that the different orientations of this token state $\ket{e_\sigma(g)}\ce U(g) \ket{e_\sigma}$ are orthogonal in the limit that $\sigma$ vanishes, $\lim_{\sigma \to 0} \braket{e_\sigma(g)|e_\sigma(g')}= \delta_{g,g'}$, imitating a classical reference frame as discussed in the previous section. In this limit token states corresponding to different positional reference frames are completely distinguishable from each other.

We must now construct the recovery measurement $R$ for which the associated set of POVM elements satisfy the covariance relation in Eq.~\eqref{covariance} with respect to the translation group $T_1$. One such set is given by the PVM elements associated with the position operator $X_R$, namely, $E(x) \ce \ket{x_R}\! \bra{x_R}$ for all $x_R\in \mathbb{R} \simeq T_1$, where $\ket{x_R}$ denotes the eigenket of $X_R$ associated with the eigenvalue $x_R$. This follows from the fact that the position and momentum operators acting on $\mathcal{H}_R$ satisfy the canonical commutation relation $[X_R, P_R]=i$, which implies that $P_R$ generates translations of the operator $X_R$, or equivalently $U_R(g) \ket{x_R} = \ket{x_R + g}$. However, there is a more general set of POVM elements corresponding to unsharp measurements of the position operator constructed by the convolution  of $E(x)$ with some confidence measure $\mu$
\begin{align}
E^{\mu}(x) \ce  \int d\mu(q) \, E(x+q). \label{}
\end{align}
Direct substitution of $E^{\mu}(x)$ into Eq.~\eqref{covariance} shows that indeed these unsharp POVM elements are covariant with respect to the translation group. In what follows we consider the family of unsharp POVM elements $E^{\mu}_{\delta}(x)$ defined by choosing a Gaussian measure parametrized by $\delta >0$,
\begin{align}
E^{\mu}_{\delta}(x) \ce  \int dq \, \frac{e^{-q^2/\delta^2}}{\sqrt{\pi} \delta} E(x+q). \label{POVMdelta}
\end{align}
In the limit $\delta \to 0$, we have $E^{\mu}_{\delta}(x) \to E(x)$.

Given that Alice prepared the reference token in the state $\rho_R = \ket{e_\sigma}\! \bra{e_\sigma} \in \mathcal{S}(\mathcal{H}_R)$, the probability distribution $p\!\left(g\right)$ appearing in Eq.~\eqref{LimitOfCompostion} is
\begin{align}
p\!\left(g\right)\ce \tr \big( E^{\mu}_{\delta}(g)  \rho_R\big) = \frac{ e^{-  \frac{g^2}{\sigma^2 + \delta^2} }}{\sqrt{\pi} \sqrt{\sigma^2 + \delta^2 }}  .
\end{align}
We note that $p\!\left(g\right)$ is peaked around $g=0$ with a width of $\sqrt{\sigma^2 + \delta^2 }$. From Eq.~\eqref{LimitOfCompostion}, and the discussion that immediately follows, we see that the parameter $\sqrt{\sigma^2 + \delta^2 }$ determines the quality of the recovery operation: the smaller $\sigma$ and $\delta$ are, the more peaked $p\!\left(g\right)$ is around the identity element and the closer Bob's recovered state will be to the state sent by Alice.

\begin{figure}[t]
\includegraphics[width =0.45\textwidth]{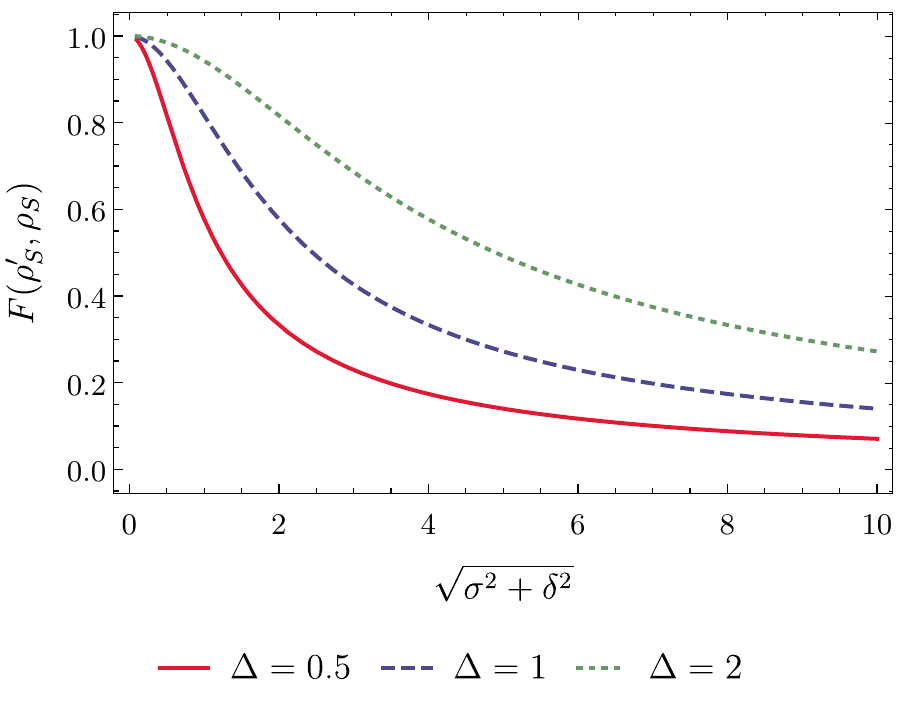}
\caption{
The fidelity $F(\rho_S',\rho_S)$ between the state sent by Alice $\rho_S$ and the state recovered by Bob $\rho_S'$ as a function of $\sqrt{\sigma^2 + \delta^2}$, where $\sigma$ is the width of the reference token in position space and $\delta$ quantifies the accuracy of Bob's measurement of the reference token. It is seen that for a fixed $\sqrt{\sigma^2 + \delta^2}$, states less localized in the position basis (larger $\Delta$) are better recovered by Bob.
 }
\label{FidelityPlot}
\end{figure}

As a concrete example, suppose Alice wishes to send Bob the state $\rho_S = \ket{\psi_S}\! \bra{\psi_S}$, where $\ket{\psi_S} \in \mathcal{H}_S \simeq L^2(\mathbb{R})$ is a Gaussian state
\begin{align}
\ket{\psi_S} &= \frac{1}{\pi^{1/4} \sqrt{\Delta}} \int dx_S \, e^{i \mu_p x} e^{-  \left(x_S-\mu_x\right)^2/2\Delta^2}  \ket{x_S}, \label{GaussSystemState}% \nn \\
%&= \frac{\sqrt{\Delta}}{\pi^{1/4}} \int dp_S \, e^{-i \mu_x p} e^{- \Delta^2 \left(p_S-\mu_p\right)^2/2}  \ket{p_S},
\end{align}
and $\Delta$ is the width of the Gaussian state in the position basis $\ket{x_S}$ for $\mathcal{H}_S$, and $\mu_x$ and $\mu_p$ are its average position and momentum. Using Eq.~\eqref{FidelityResult}, the fidelity between $\rho_S$ and the state recovered by Bob $\rho_S'$ is
\begin{align}
F(\rho_S',\rho_S) %&= \int dp_S dp_S' \, \tilde{p}(p_S - p_S')  \abs{\psi_S(p_S)}^2 \abs{\psi_S(p_S')}^2 \nn \\
%&= \int dp_S dp_S' \, \frac{e^{-\sigma^2 (p_S-p_S')^2/4 }}{\sqrt{2\pi}}  \left( \frac{\Delta }{\sqrt{\pi}}e^{-\Delta^2 p_s^2}\right) \left( \frac{\Delta }{\sqrt{\pi}}e^{-\Delta^2 p_s'^2}\right) \nn \\
&= \frac{\Delta}{\sqrt{ \Delta^2 + \frac{1}{2} \left(\sigma^2 + \delta^2\right)}}. \label{FidelityTranslation}
\end{align}
As might be expected, in the limit where $\sigma$ and $\delta$ vanish the fidelity $F(\rho_S',\rho_S)$ is equal to unity and the recovered state is exactly equal to the state Alice wished to send to Bob, $\rho_S' = \rho_S$. This limit corresponds different orientations of the reference token described by Eq.~\eqref{SetOfReferenceTokens} being orthogonal, thus imitating a classical reference frame, and the measurement of the token's position being carried out perfectly.

From Eq.~\eqref{FidelityTranslation} we also observe that states less localized in the position basis (larger~$\Delta$) are better recovered by Bob, as illustrated in Fig.~\ref{FidelityPlot} in which the fidelity is plotted as a function of $\sqrt{\sigma^2 + \delta^2 }$ for different $\Delta$. Note that the expression for the fidelity is independent of $\mu_x$ and $\mu_p$, implying that for Gaussian states the success of the recovery operation is independent of where the state is localized in phase space.

As a second example, suppose Alice prepares her token in a superposition of two Gaussian wave packets
\begin{align}
\ket{e} = \frac{1}{\sqrt{N}} \big( \ket{\psi(\bar{x},\bar{p},\sigma)} + \ket{\psi(-\bar{x},-\bar{p},\sigma)} \big) \in \mathcal{H}_R, 
\label{GaussSuper}
\end{align}
where $N$ is an appropriate normalization constant and $\ket{\psi(\bar{x},\bar{p},\sigma)}$ denotes the state of a Gaussian wave packet of width $\sigma$ in position space with average position and momentum $\bar{x}$ and $\bar{p}$, respectively. As they appear in Eq.~\eqref{GaussSuper}, $\bar{x}$ and $\bar{p}$ quantify the size of the superposition in position and momentum space, respectively. Further, suppose that Bob is able to make a perfect measurement of the position of the reference token as described by the POVM elements $\lim_{\delta \to 0} E^{\mu}_{\delta}(x)$. And again, suppose Alice  wishes to communicate the Gaussian state given in Eq.~\eqref{GaussSystemState}. 

\begin{figure}[t]
\includegraphics[width =0.45\textwidth]{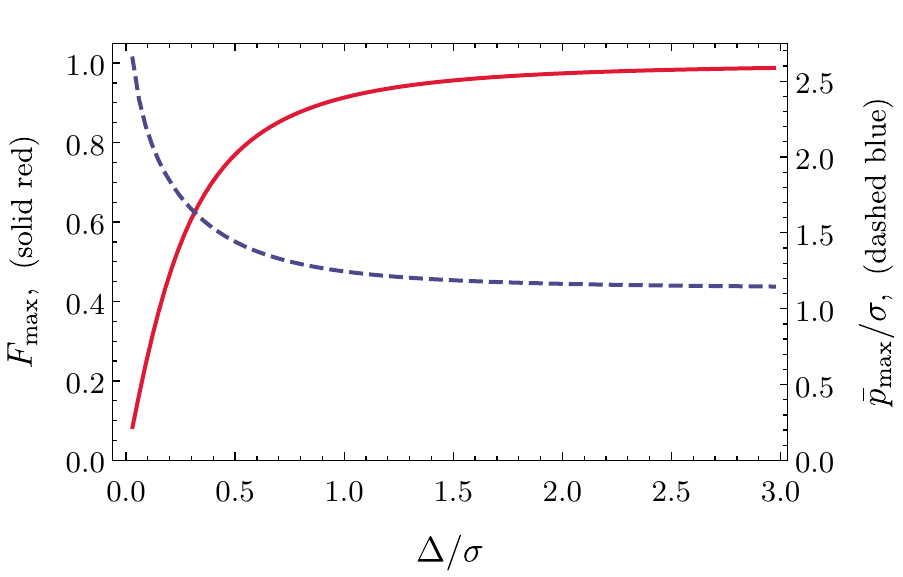}
\caption{
For a reference token prepared in a superposition of two Gaussian states described by Eq.~\eqref{GaussSuper}, the maximum fidelity $F_{\rm max} \ce \max \left[F(\rho_S',\rho_S)\ | \ \bar{x}, \bar{p}, \sigma >0 \right]$ and the size of the superposition in momentum space $\bar{p}_{\rm max}/\sigma$ that realizes this maximum is plotted as a function of the width of the in position space $\Delta/\sigma$ of the state Alice sent to Bob as given in Eq.~\eqref{GaussSystemState}. For all values of $\Delta/\sigma$ the size of the superposition in momentum space which realizes the maximum fidelity is~$\bar{x}_{\rm max} \sigma = 0$.}
\label{plot2}
\end{figure}

Given the above, the fidelity expressed in Eq.~\eqref{FidelityResult} yields
\begin{align}
F(\rho_S',\rho_S) &= \beta \, \frac{e^{ \beta^2  \bar{x}^2 /\sigma^2  } 
+e^{ -\beta^2 \bar{p}^2 \sigma^2 }
}{e^{\bar{x}^2 / \sigma^2 } + e^{ -\bar{p}^2  \sigma^2 }},\label{supFidelity}
\end{align}
where $\beta \ce \Delta/ \sqrt{\Delta^2 + \sigma^2/2}$; note that $\beta \in (0,1)$ and is equal to Eq.~\eqref{FidelityTranslation} when $\delta \to 0$. Further, $\beta$ takes its maximum (minimum) value when $\Delta \gg \sigma$ ($\Delta \ll \sigma$). Observe that the fidelity in Eq.~\eqref{supFidelity} is independent of $\mu_x$ and $\mu_p$ appearing in Eq.~\eqref{GaussSystemState}, implying that the success of the recovery operation is independent of where $\ket{\psi_S}$ is localized in phase space.

The fidelity in Eq.~\eqref{supFidelity} is a monotonically decreasing function of $\bar{x}$, which implies that Alice should prepare the size of the superposition in position space to be as small as possible (i.e., small $\bar{x}$) in order to maximize the fidelity. A second observation can be made by inspection of Fig.~\ref{plot2}, which is a plot of both the maximum fidelity, $F_{\rm max} \ce \max \left[F(\rho_S',\rho_S)\ | \ \bar{x}, \bar{p}, \sigma >0 \right]$, and the value $\bar{p}_{\rm max}/\sigma$ which realizes this maximum as a function of the width $\Delta/\sigma$ of the state $\ket{\psi_S}$ Alice wishes to send to Bob; since the fidelity is monotonically decreasing in  $\bar{x} \sigma$, this maximum occurs when $\bar{x} \sigma=0$. From Fig.~\ref{plot2} we see that depending on the value of $\Delta/\sigma$, Alice can adjust the state of the reference token by choosing the size of the superposition in momentum space $\bar{p}/\sigma$ so that the fidelity is maximized. That is, having the ability to create different sizes of superposition in momentum space can act as a resource to improve the communication channel specific to the state Alice wishes to send to Bob.

\section{Conclusions and Outlook}
\label{SummaryCh7}

We began by introducing a communication protocol between two parties, Alice and Bob, that do not share a reference frame associated with a compact group. Alice sends to Bob a token of her reference frame along with a system she wishes to communicate to him, and then Bob performs an appropriate recovery operation that enables him to recover a state close to the one Alice wished to communicate. 

In Sec.~\ref{A recovery operation for noncompact groups} we showed that this communication protocol can be applied when Alice's and Bob's reference frames are associated with a noncompact group, even though averaging states over the entire group leads to non-normalizable states. Furthermore, we demonstrated that this communication channel induces decoherence in the basis furnished by the eigenkets of the generator of the group.  In Sec.~\ref{Application to reference frames with the translation group} we applied this result to the study of communication between two parties who do not share a reference frame associated with the translation group. We introduced a sequence of Gaussian states $\ket{e_\sigma}$ of the reference token with spatial width $\sigma$, and saw that in the limit $\sigma \to 0$, $\ket{e_\sigma}$ imitates a classical reference frame. This suggests that the parameter $1/\sigma$ acts as the effective size of the reference token, since as $1/\sigma$ becomes large the two parties are able to communicate perfectly (assuming Bob is able to measure the reference token perfectly, $\delta \to 0$). We also demonstrated that for finite size reference tokens, i.e., when $1/\sigma$ is finite, states less localized in the position basis are better communicated to Bob and examined the case when the reference token is prepared in a superposition. 

We note that the group of time translations generated by a Hamiltonian is a strongly continuous one-dimensional noncompact Lie group. Thus, provided a covariant measurement of the reference token corresponding to a time observable can be constructed \cite{Busch:1997}, the above communication scheme can be employed. This will be fruitful for communication between parties who do not share a temporal reference frame, that is, their clocks are not synchronized. Furthermore, it will be interesting to see how the above construction can be applied to noncompact Lie groups of higher dimension, such as the Euclidean group in two and three dimensions, the Galilean group, and ultimately the Poincar\'{e} group. 

The intended application of the results in this article, as well as one of the primary motivation for this investigation, is to study the act of changing quantum reference frames\footnote{See Refs. \cite{Giacomini:2018, Vanrietvelde:2018, Hoehn:2018} for a different approach.}. Palmer \emph{et al.} \cite{Palmer:2013} have constructed an operational protocol for changing quantum reference frames associated with compact groups. They used the state $\mathcal{G}[\rho_A \otimes \rho_S]$ as a relational description of the state $\rho_S$ with respect to a quantum reference frame $\rho_A$, and then considered the operation of changing the quantum reference frame from the state $\rho_A$ to $\rho_B$. They found that this operation could not be done perfectly and that the best one could do is
 \begin{align}
\mathcal{G}[\rho_A \otimes \rho_S] \to \mathcal{G}[\rho_B \otimes \rho_S'],
\end{align}
where $\rho_S' = \mathcal{R} \circ \mathcal{E} [\rho_S]$. In other words, one is not able to change quantum reference frames without affecting the state of the system described with respect to the reference frame\,---\,$\rho_S$ changes to $\rho_S'$ when the reference frame is changed. This results in a fundamental decoherence mechanism associated with the act of changing quantum reference frames. 
This is in stark contrast to the classical case, in which the act of changing reference frames does not affect the system being described with respect to the reference frames. This decoherence is described by the composition of the encoding and recovery operations $\mathcal{R} \circ \mathcal{E}$ discussed in this article. Having generalized the operation $\mathcal{R} \circ \mathcal{E}$ to reference frames associated with noncompact groups, we hope to study the effect of changing quantum reference frames associated with the  Galilean  and Poincar\'{e} groups. Understanding the process of changing quantum reference frames is an essential step in the construction of a relational quantum theory, in which all objects, including reference frames, are treated quantum mechanically. 

\begin{acknowledgments}
I wish to thank Marco Piani, Robert B. Mann, and Urgje~\cite{219979} for useful discussions and Mehdi Ahmadi and Lorenza Viola for a careful reading of this manuscript. This work was supported by the Natural Sciences and Engineering Research Council of Canada and the Dartmouth College Society of Fellows.
\end{acknowledgments}

\bibliography{CommunicationRef}

%merlin.mbs apsrev4-1.bst 2010-07-25 4.21a (PWD, AO, DPC) hacked
%Control: key (0)
%Control: author (8) initials jnrlst
%Control: editor formatted (1) identically to author
%Control: production of article title (-1) disabled
%Control: page (0) single
%Control: year (1) truncated
%Control: production of eprint (0) enabled
\begin{thebibliography}{20}%
\makeatletter
\providecommand \@ifxundefined [1]{%
 \@ifx{#1\undefined}
}%
\providecommand \@ifnum [1]{%
 \ifnum #1\expandafter \@firstoftwo
 \else \expandafter \@secondoftwo
 \fi
}%
\providecommand \@ifx [1]{%
 \ifx #1\expandafter \@firstoftwo
 \else \expandafter \@secondoftwo
 \fi
}%
\providecommand \natexlab [1]{#1}%
\providecommand \enquote  [1]{``#1''}%
\providecommand \bibnamefont  [1]{#1}%
\providecommand \bibfnamefont [1]{#1}%
\providecommand \citenamefont [1]{#1}%
\providecommand \href@noop [0]{\@secondoftwo}%
\providecommand \href [0]{\begingroup \@sanitize@url \@href}%
\providecommand \@href[1]{\@@startlink{#1}\@@href}%
\providecommand \@@href[1]{\endgroup#1\@@endlink}%
\providecommand \@sanitize@url [0]{\catcode `\\12\catcode `\$12\catcode
  `\&12\catcode `\#12\catcode `\^12\catcode `\_12\catcode `\%12\relax}%
\providecommand \@@startlink[1]{}%
\providecommand \@@endlink[0]{}%
\providecommand \url  [0]{\begingroup\@sanitize@url \@url }%
\providecommand \@url [1]{\endgroup\@href {#1}{\urlprefix }}%
\providecommand \urlprefix  [0]{URL }%
\providecommand \Eprint [0]{\href }%
\providecommand \doibase [0]{http://dx.doi.org/}%
\providecommand \selectlanguage [0]{\@gobble}%
\providecommand \bibinfo  [0]{\@secondoftwo}%
\providecommand \bibfield  [0]{\@secondoftwo}%
\providecommand \translation [1]{[#1]}%
\providecommand \BibitemOpen [0]{}%
\providecommand \bibitemStop [0]{}%
\providecommand \bibitemNoStop [0]{.\EOS\space}%
\providecommand \EOS [0]{\spacefactor3000\relax}%
\providecommand \BibitemShut  [1]{\csname bibitem#1\endcsname}%
\let\auto@bib@innerbib\@empty
%</preamble>
\bibitem [{\citenamefont {Nielsen}\ and\ \citenamefont
  {Chuang}(2010)}]{Nielsen:2010}%
  \BibitemOpen
  \bibfield  {author} {\bibinfo {author} {\bibfnamefont {M.~A.}\ \bibnamefont
  {Nielsen}}\ and\ \bibinfo {author} {\bibfnamefont {I.~L.}\ \bibnamefont
  {Chuang}},\ }\href@noop {} {\emph {\bibinfo {title} {Quantum Computation and
  Quantum Information}}}\ (\bibinfo  {publisher} {Cambridge University Press},\
  \bibinfo {address} {Cambridge},\ \bibinfo {year} {2010})\BibitemShut
  {NoStop}%
\bibitem [{\citenamefont {Chiribella}\ \emph {et~al.}(2012)\citenamefont
  {Chiribella}, \citenamefont {Giovannetti}, \citenamefont {Maccone},\ and\
  \citenamefont {Perinotti}}]{Giulio-Chiribella:2012}%
  \BibitemOpen
  \bibfield  {author} {\bibinfo {author} {\bibfnamefont {G.}~\bibnamefont
  {Chiribella}}, \bibinfo {author} {\bibfnamefont {V.}~\bibnamefont
  {Giovannetti}}, \bibinfo {author} {\bibfnamefont {L.}~\bibnamefont
  {Maccone}}, \ and\ \bibinfo {author} {\bibfnamefont {P.}~\bibnamefont
  {Perinotti}},\ }\href {\doibase 10.1103/PhysRevA.86.010304} {\bibfield
  {journal} {\bibinfo  {journal} {Phys. Rev. A}\ }\textbf {\bibinfo {volume}
  {86}},\ \bibinfo {pages} {010304(R)} (\bibinfo {year} {2012})}\BibitemShut
  {NoStop}%
\bibitem [{\citenamefont {Verdon}\ and\ \citenamefont
  {Vicary}(2018{\natexlab{a}})}]{Verdon:2018}%
  \BibitemOpen
  \bibfield  {author} {\bibinfo {author} {\bibfnamefont {D.}~\bibnamefont
  {Verdon}}\ and\ \bibinfo {author} {\bibfnamefont {J.}~\bibnamefont
  {Vicary}},\ }\href {\doibase 10.1103/PhysRevA.98.012306} {\bibfield
  {journal} {\bibinfo  {journal} {Phys. Rev. A}\ }\textbf {\bibinfo {volume}
  {98}},\ \bibinfo {pages} {012306} (\bibinfo {year}
  {2018}{\natexlab{a}})}\BibitemShut {NoStop}%
\bibitem [{\citenamefont {Verdon}\ and\ \citenamefont
  {Vicary}(2018{\natexlab{b}})}]{Verdon2:2018}%
  \BibitemOpen
  \bibfield  {author} {\bibinfo {author} {\bibfnamefont {D.}~\bibnamefont
  {Verdon}}\ and\ \bibinfo {author} {\bibfnamefont {J.}~\bibnamefont
  {Vicary}},\ }\href {https://arxiv.org/abs/1802.09040} {\bibfield  {journal}
  {\bibinfo  {journal} {arXiv:1802.09040v2 [quant-ph]}\ } (\bibinfo {year}
  {2018}{\natexlab{b}})}\BibitemShut {NoStop}%
\bibitem [{\citenamefont {Bartlett}\ \emph {et~al.}(2007)\citenamefont
  {Bartlett}, \citenamefont {Rudolph},\ and\ \citenamefont
  {Spekkens}}]{Bartlett:2007}%
  \BibitemOpen
  \bibfield  {author} {\bibinfo {author} {\bibfnamefont {S.~D.}\ \bibnamefont
  {Bartlett}}, \bibinfo {author} {\bibfnamefont {T.}~\bibnamefont {Rudolph}}, \
  and\ \bibinfo {author} {\bibfnamefont {R.~W.}\ \bibnamefont {Spekkens}},\
  }\href {\doibase 10.1103/RevModPhys.79.555} {\bibfield  {journal} {\bibinfo
  {journal} {Rev. Mod. Phys.}\ }\textbf {\bibinfo {volume} {79}},\ \bibinfo
  {pages} {555} (\bibinfo {year} {2007})}\BibitemShut {NoStop}%
\bibitem [{\citenamefont {Bartlett}\ \emph {et~al.}(2003)\citenamefont
  {Bartlett}, \citenamefont {Rudolph},\ and\ \citenamefont
  {Spekkens}}]{Rudolph:2003}%
  \BibitemOpen
  \bibfield  {author} {\bibinfo {author} {\bibfnamefont {S.~D.}\ \bibnamefont
  {Bartlett}}, \bibinfo {author} {\bibfnamefont {T.}~\bibnamefont {Rudolph}}, \
  and\ \bibinfo {author} {\bibfnamefont {R.~W.}\ \bibnamefont {Spekkens}},\
  }\href {\doibase 10.1103/PhysRevLett.91.027901} {\bibfield  {journal}
  {\bibinfo  {journal} {Phys. Rev. Lett.}\ }\textbf {\bibinfo {volume} {91}},\
  \bibinfo {pages} {027901} (\bibinfo {year} {2003})}\BibitemShut {NoStop}%
\bibitem [{\citenamefont {Bartlett}\ \emph {et~al.}(2009)\citenamefont
  {Bartlett}, \citenamefont {Rudolph}, \citenamefont {Spekkens},\ and\
  \citenamefont {Turner}}]{Bartlett:2009}%
  \BibitemOpen
  \bibfield  {author} {\bibinfo {author} {\bibfnamefont {S.~D.}\ \bibnamefont
  {Bartlett}}, \bibinfo {author} {\bibfnamefont {T.}~\bibnamefont {Rudolph}},
  \bibinfo {author} {\bibfnamefont {R.~W.}\ \bibnamefont {Spekkens}}, \ and\
  \bibinfo {author} {\bibfnamefont {P.~S.}\ \bibnamefont {Turner}},\ }\href
  {\doibase 10.1088/1367-2630/11/6/063013} {\bibfield  {journal} {\bibinfo
  {journal} {New J. Phys.}\ }\textbf {\bibinfo {volume} {11}},\ \bibinfo
  {pages} {063013} (\bibinfo {year} {2009})}\BibitemShut {NoStop}%
\bibitem [{\citenamefont {Smith}\ \emph {et~al.}(2016)\citenamefont {Smith},
  \citenamefont {Piani},\ and\ \citenamefont {Mann}}]{Smith:2016}%
  \BibitemOpen
  \bibfield  {author} {\bibinfo {author} {\bibfnamefont {A.~R.~H.}\
  \bibnamefont {Smith}}, \bibinfo {author} {\bibfnamefont {M.}~\bibnamefont
  {Piani}}, \ and\ \bibinfo {author} {\bibfnamefont {R.~B.}\ \bibnamefont
  {Mann}},\ }\href {\doibase 10.1103/PhysRevA.94.012333} {\bibfield  {journal}
  {\bibinfo  {journal} {Phys. Rev. A}\ }\textbf {\bibinfo {volume} {94}},\
  \bibinfo {pages} {012333} (\bibinfo {year} {2016})}\BibitemShut {NoStop}%
\bibitem [{\citenamefont {Miatto}(2012)}]{Miatto:2012}%
  \BibitemOpen
  \bibfield  {author} {\bibinfo {author} {\bibfnamefont {F.~M.}\ \bibnamefont
  {Miatto}},\ }\href {https://arxiv.org/abs/1209.4281} {\bibfield  {journal}
  {\bibinfo  {journal} {arXiv:1209.4281 [quant-ph]}\ } (\bibinfo {year}
  {2012})}\BibitemShut {NoStop}%
\bibitem [{\citenamefont {Ahmadi}\ \emph {et~al.}(2015)\citenamefont {Ahmadi},
  \citenamefont {Smith},\ and\ \citenamefont {Dragan}}]{Ahmadi:2015}%
  \BibitemOpen
  \bibfield  {author} {\bibinfo {author} {\bibfnamefont {M.}~\bibnamefont
  {Ahmadi}}, \bibinfo {author} {\bibfnamefont {A.~R.~H.}\ \bibnamefont
  {Smith}}, \ and\ \bibinfo {author} {\bibfnamefont {A.}~\bibnamefont
  {Dragan}},\ }\href {\doibase 10.1103/PhysRevA.92.062319} {\bibfield
  {journal} {\bibinfo  {journal} {Phys. Rev. A}\ }\textbf {\bibinfo {volume}
  {92}},\ \bibinfo {pages} {062319} (\bibinfo {year} {2015})}\BibitemShut
  {NoStop}%
\bibitem [{\citenamefont {Nachbin}(1965)}]{Nachbin:1965}%
  \BibitemOpen
  \bibfield  {author} {\bibinfo {author} {\bibfnamefont {L.}~\bibnamefont
  {Nachbin}},\ }\href@noop {} {\emph {\bibinfo {title} {The Haar Integral}}}\
  (\bibinfo  {publisher} {D. Van Nostrand Company,},\ \bibinfo {address}
  {Princeton, N.J.},\ \bibinfo {year} {1965})\BibitemShut {NoStop}%
\bibitem [{\citenamefont {Stone}(1930)}]{Stone:1930}%
  \BibitemOpen
  \bibfield  {author} {\bibinfo {author} {\bibfnamefont {M.~H.}\ \bibnamefont
  {Stone}},\ }\href {\doibase 10.1073/pnas.16.2.172} {\bibfield  {journal}
  {\bibinfo  {journal} {PNAS}\ }\textbf {\bibinfo {volume} {16}},\ \bibinfo
  {pages} {172} (\bibinfo {year} {1930})}\BibitemShut {NoStop}%
\bibitem [{\citenamefont {Ballentine}(1998)}]{Ballentine:1998}%
  \BibitemOpen
  \bibfield  {author} {\bibinfo {author} {\bibfnamefont {L.~E.}\ \bibnamefont
  {Ballentine}},\ }\href@noop {} {\emph {\bibinfo {title} {Quantum Mechanics: A
  Modern Development}}}\ (\bibinfo  {publisher} {World Scientific},\ \bibinfo
  {year} {1998})\BibitemShut {NoStop}%
\bibitem [{\citenamefont {Busch}\ \emph {et~al.}(1997)\citenamefont {Busch},
  \citenamefont {Grabowski},\ and\ \citenamefont {Lahti}}]{Busch:1997}%
  \BibitemOpen
  \bibfield  {author} {\bibinfo {author} {\bibfnamefont {P.}~\bibnamefont
  {Busch}}, \bibinfo {author} {\bibfnamefont {M.}~\bibnamefont {Grabowski}}, \
  and\ \bibinfo {author} {\bibfnamefont {P.~J.}\ \bibnamefont {Lahti}},\
  }\href@noop {} {\emph {\bibinfo {title} {Operational Quantum Physics}}}\
  (\bibinfo  {publisher} {Springer},\ \bibinfo {year} {1997})\BibitemShut
  {NoStop}%
\bibitem [{\citenamefont {Rudin}(1976)}]{Rudin:1976}%
  \BibitemOpen
  \bibfield  {author} {\bibinfo {author} {\bibfnamefont {W.}~\bibnamefont
  {Rudin}},\ }\href@noop {} {\emph {\bibinfo {title} {Principles of
  mathematical analysis}}}\ (\bibinfo  {publisher} {McGraw-Hill},\ \bibinfo
  {address} {New York},\ \bibinfo {year} {1976})\BibitemShut {NoStop}%
\bibitem [{\citenamefont {Giacomini}\ \emph {et~al.}(2018)\citenamefont
  {Giacomini}, \citenamefont {Castro-Ruiz},\ and\ \citenamefont {\u{C}aslav
  Brukner}}]{Giacomini:2018}%
  \BibitemOpen
  \bibfield  {author} {\bibinfo {author} {\bibfnamefont {F.}~\bibnamefont
  {Giacomini}}, \bibinfo {author} {\bibfnamefont {E.}~\bibnamefont
  {Castro-Ruiz}}, \ and\ \bibinfo {author} {\bibnamefont {\u{C}aslav
  Brukner}},\ }\href {https://arxiv.org/abs/1712.07207} {\bibfield  {journal}
  {\bibinfo  {journal} {arXiv:1712.07207 [quant-ph]}\ } (\bibinfo {year}
  {2018})}\BibitemShut {NoStop}%
\bibitem [{\citenamefont {Vanrietvelde}\ \emph {et~al.}(2018)\citenamefont
  {Vanrietvelde}, \citenamefont {Hoehn}, \citenamefont {Giacomini},\ and\
  \citenamefont {Castro-Ruiz}}]{Vanrietvelde:2018}%
  \BibitemOpen
  \bibfield  {author} {\bibinfo {author} {\bibfnamefont {A.}~\bibnamefont
  {Vanrietvelde}}, \bibinfo {author} {\bibfnamefont {P.~A.}\ \bibnamefont
  {Hoehn}}, \bibinfo {author} {\bibfnamefont {F.}~\bibnamefont {Giacomini}}, \
  and\ \bibinfo {author} {\bibfnamefont {E.}~\bibnamefont {Castro-Ruiz}},\
  }\href {https://arxiv.org/abs/1809.00556} {\bibfield  {journal} {\bibinfo
  {journal} {arXiv:1809.00556 [quant-ph]}\ } (\bibinfo {year}
  {2018})}\BibitemShut {NoStop}%
\bibitem [{\citenamefont {Hoehn}\ and\ \citenamefont
  {Vanrietvelde}(2018)}]{Hoehn:2018}%
  \BibitemOpen
  \bibfield  {author} {\bibinfo {author} {\bibfnamefont {P.~A.}\ \bibnamefont
  {Hoehn}}\ and\ \bibinfo {author} {\bibfnamefont {A.}~\bibnamefont
  {Vanrietvelde}},\ }\href {https://arxiv.org/abs/1810.04153} {\bibfield
  {journal} {\bibinfo  {journal} {arXiv:1810.04153 [gr-qc]}\ } (\bibinfo {year}
  {2018})}\BibitemShut {NoStop}%
\bibitem [{\citenamefont {Palmer}\ \emph {et~al.}(2014)\citenamefont {Palmer},
  \citenamefont {Girelli},\ and\ \citenamefont {Bartlett}}]{Palmer:2013}%
  \BibitemOpen
  \bibfield  {author} {\bibinfo {author} {\bibfnamefont {M.~C.}\ \bibnamefont
  {Palmer}}, \bibinfo {author} {\bibfnamefont {F.}~\bibnamefont {Girelli}}, \
  and\ \bibinfo {author} {\bibfnamefont {S.~D.}\ \bibnamefont {Bartlett}},\
  }\href {\doibase 10.1103/PhysRevA.89.052121} {\bibfield  {journal} {\bibinfo
  {journal} {Phys. Rev. A}\ }\textbf {\bibinfo {volume} {89}},\ \bibinfo
  {pages} {052121} (\bibinfo {year} {2014})}\BibitemShut {NoStop}%
\bibitem [{\citenamefont {Urgje}(2015)}]{219979}%
  \BibitemOpen
  \bibfield  {author} {\bibinfo {author} {\bibnamefont {Urgje}},\ }\href
  {https://physics.stackexchange.com/q/219979} {\bibfield  {journal} {\bibinfo
  {journal} {https://physics.stackexchange.com/q/219979}\ } (\bibinfo {year}
  {2015})}\BibitemShut {NoStop}%
\end{thebibliography}%
\end{document}